\documentclass[11pt]{article}
\usepackage[utf8]{inputenc}
\usepackage[english]{babel}
\usepackage{amsmath, amsfonts, amssymb}
\usepackage{nccmath}
\usepackage{mathrsfs}
\usepackage[mathscr]{euscript}
\usepackage{color}
\usepackage{amsthm}
\usepackage{psfrag}
\usepackage[toc,page]{appendix}
\usepackage{graphicx}
\usepackage{url}
\usepackage[hyperfootnotes=false]{hyperref}
\usepackage{multirow}
\usepackage{makecell}
\usepackage{bbm}
\usepackage{comment}
\usepackage{float}
\usepackage[normalem]{ulem}
\usepackage[backend=biber, sorting=none]{biblatex}
\catcode`@=11
\@addtoreset{equation}{section}
\catcode`@=12

\newcommand{\be}{\begin{equation}}
\newcommand{\ee}{\end{equation}}
\newcommand{\bl}{\begin{align}}
\newcommand{\el}{\end{align}}
\newcommand{\ba}{\begin{aligned}}
\newcommand{\ea}{\end{aligned}}
\newcommand{\beqa}{\begin{eqnarray}}
\newcommand{\eeqa}{\end{eqnarray}}

\renewcommand\l{\lambda}

\def\e{{\rm e}}

\newcommand{\bseq}{\begin{subequations}}
\newcommand{\eseq}{\end{subequations}}

\renewcommand{\tanh}{\mathop{\rm th}\nolimits}

\numberwithin{equation}{section}  

\begin{document}
\begin{titlepage}
\clearpage

\title{{\bf Space- vs Time-dependence in taming the infrared instability of projectable Ho\v rava Gravity} }
\author{Shinji Mukohyama$^{a,b,c}$\footnote{shinji.mukohyama@yukawa.kyoto-u.ac.jp}, Jury Radkovski$^{d,e}$\footnote{jradkovski@perimeterinstitute.ca},  \, Sergey Sibiryakov$^{d,e}$\footnote{ssibiryakov@perimeterinstitute.ca}~\\[2mm]
{\small \it $^a$Center for Gravitational Physics and Quantum Information,} \\
{\small \it Yukawa Institute for Theoretical Physics, Kyoto University, 606-8502 Kyoto, Japan}\\
{\small \it $^b$Research Center for the Early Universe (RESCEU), } \\
{\small \it Graduate School of Science, The University of Tokyo,}\\
{\small \it  Hongo 7-3-1, Bunkyo-ku, Tokyo 113-0033, Japan}\\
{\small \it $^c$ Kavli Institute for the Physics and Mathematics of the Universe (WPI),}\\
{\small \it The University of Tokyo Institutes for Advanced Study (UTIAS),}\\
{\small \it The University of Tokyo, Kashiwa, Chiba 277-8583, Japan} \\
{\small \it $^d$Department of Physics and Astronomy, McMaster University,}\\
{\small \it 1280 Main St W, Hamilton, ON L8S 4M1, Canada}\\ 
{\small\it $^e$Perimeter Institute for Theoretical Physics, 31 Caroline St N,  Waterloo, ON N2L 2Y5, Canada}
}
\date{}

\maketitle

\begin{abstract}
Minkowski spacetime
exhibits infrared instability 
in projectable Ho\v rava gravity in $(3+1)$ dimensions.
To be phenomenologically viable, the instability should be either hidden by other time-dependent processes such as the Hubble expansion of the universe and the Jeans instability, or evolve into another static solution with low average curvature.
While the former scenario leads to a phenomenological constraint on the infrared properties of the renormalization group flow, this paper explores the latter possibility. 
We study if the presence of higher derivative terms in the action can lead to existence of 
static, inhomogeneous (qua\-si-) periodic solutions with planar symmetry, similar to modulated phases in magnetic materials. We find that such solutions do not exist. In doing so, we classify all static homogeneous and isotropic solutions and solutions with planar symmetry. We provide arguments that none of them can serve as an endpoint for the evolution of the Minkowski instability. This motivates further study of the scenario where the instability is concealed by time evolution. 
\end{abstract} 

\begin{flushright} YITP-26-35, RESCEU-10/26, IPMU26-0013 \end{flushright}

\thispagestyle{empty}
\end{titlepage}

\newpage

\tableofcontents

\section{Introduction}
\label{eq:introduction}
Ho\v rava gravity (HG) \cite{Horava2008, Horava2009} is a quantum theory of gravity formulated as a unitary renormalizable gauge field theory (see the reviews \cite{Mukohyama2010, Sotiriou2010, Blas:2014aca, Barvinsky2023Review,Herrero-Valea2023}). It is defined by a path integral over metrics corresponding to spacetimes with topological structure in the form of a spacelike \textit{foliation}. The foliation admits the gauge symmetry group of \textit{foliation preserving diffeomorphisms} (FDiffs)
\begin{subequations}
\label{eq:FDiffsCoords}
\begin{equation}
\label{eq:spatialdiffs}
    \mathbf{x} \mapsto \tilde{\mathbf{x}}(t, \mathbf{x}) \, ,  \\ 
\end{equation}
\begin{equation}
\label{eq:timereparam}
t \mapsto
    \tilde{t}(t) \, .
\end{equation}
\end{subequations}
The time reparameterizations here are \textit{projectable} functions, that is, functions depending on a foliation leaf as a whole, and so \eqref{eq:FDiffsCoords} is a subgroup of the full spacetime diffeomorphisms.

The path integral of HG is further assumed to be dominated in the ultraviolet (UV) by an asymptotically free fixed point with anisotropic scaling of space and time,
\begin{equation}
\label{AnisotropicScaling}
    {\bf x} \rightarrow b^{-1} \, {\bf x}, \quad t \rightarrow b^{-z} \,
    t \, , 
\end{equation}
with dynamical \textit{critical exponent} $z$, equal to the dimension of space $d$ at high energies. The combination of the foliation structure plus anisotropic scaling allows us to write the action for the metric with higher spatial derivatives and with no extra time derivatives. This improves the convergence of the loop integrals without bringing any ghost degrees of freedom, unlike the fully generally covariant higher-derivative gravity \cite{Stelle1976, Stelle1977}.\footnote{See also \cite{Aoki:2019snr} for an attempt towards ghost-free covariant quadratic gravity with dynamical torsion and \cite{Mukohyama:2013gra} for a renormalizable scalar-tensor theory with Euclidean signature.}

The theory is naturally described in terms of the Arnowitt--Deser--Misner (ADM) variables adapted to the foliation \cite{Arnowitt59}. The $(d+1)$-dimensional metric is represented as 
\begin{align}
\label{Interval}
    ds^2 = -N^2 dt^2 + \gamma_{ij} (dx^i + N^i dt)(dx^j + N^j dt), \quad i = 1, \dots ,d \, \, ,
\end{align}
with lapse function $N$, shift vector $N^i$, and the spatial metric on the leaves $\gamma_{ij}$. 
The most general action consistent with the FDiff symmetry \eqref{eq:FDiffsCoords} reads \cite{Horava2009, Blas2009},
\begin{equation}
\label{eq:HG_Action}
    S = \frac{1}{2 G} \int d^3 x dt N
\sqrt{\gamma} \left( K_{ij}K^{ij} - \lambda K^2 - \mathcal{V}[\nabla_i \log N, R_{ijkl}] \right) \, ,
\end{equation}
where $K_{ij}$ is the extrinsic curvature of the foliation leaves,
\begin{equation}
\label{eq:Kij}
    K_{ij} = \frac{1}{2N}\big(
\dot{\gamma}_{ij} -\nabla_{i}N_{j}-\nabla_{j}N_{i}\big)\;,
\end{equation}
and the potential $\mathcal{V}$ is built out of the FDiff-invariant combination $\nabla_i \log N$ and curvature invariants constructed from the spatial metric $\gamma_{ij}$ using its Riemann tensor $R_{ijkl}$, Ricci tensor $R_{jl}=\gamma^{ik} R_{ijkl}$, and scalar curvature $R=\gamma^{ij}R_{ij}$.
Apart from the gravitational coupling $G$ the action contains a new coupling $\lambda$, as well as various couplings parameterizing the operators in the potential $\mathcal{V}$. The full, \textit{non-projectable} HG, after a proper parameter tuning, is a phenomenologically viable theory of gravity \cite{EmirGumrukcuoglu2017}. Its consistency as a quantum theory is however still a matter of research. The potential has $\mathcal{O}(100)$ couplings and the theory exhibits mixed first and second-class constraint structure \cite{Donnelly2011}. Together the two features complicate the analysis of its renormalizability. See \cite{Bellorin2023, Blas2025} for a recent progress
towards establishing its UV properties.

In this work we focus on simpler\footnote{The number of couplings is of $\mathcal{O}(10)$ and all constraints are first-class.} \textit{projectable} version of HG where the lapse is taken to be a function of time only, consistently with the FDiffs \eqref{eq:FDiffsCoords},
\begin{equation}
\label{eq:Projectability}
    N = N(t) \, .
\end{equation}
Using the transformations \eqref{eq:timereparam} we can always gauge fix 
\begin{equation}
\label{eq:Projectability2}
    N=1 \, , 
\end{equation}
and work with the theory with only spatial gauge diffeomorphisms \eqref{eq:spatialdiffs}. Locally, the two formulations \eqref{eq:Projectability}, \eqref{eq:Projectability2} are equivalent. Globally, the difference amounts to the existence of a global Hamiltonian constraint, associated to the gauge symmetry \eqref{eq:timereparam}. 

In this work we fix the number of space dimensions to the phenomenologically relevant case,
\begin{equation}
    d=3 \, .
\end{equation}
The potential then assumes a relatively simple form,
\begin{align}
\label{eq:potential}
    \mathcal{V} &= 2 \Lambda - \eta R +  \mu_1 R^2 + \mu_2 R_{i j } R^{i j} \nonumber \\ 
    &+ \nu_1 R^3 + \nu_2 R R_{ij} R^{ij} + \nu_3 R^{i}_{j} R^{j}_{k}R^{k}_{i} + \nu_4 \nabla_i R \nabla^i R + \nu_5 \nabla_i R_{jk} \nabla^{i} R^{jk} \, .
\end{align}
Note that by setting $\lambda = \eta = 1$ and $\mu_i = \nu_i = 0$ in \eqref{eq:HG_Action}, \eqref{eq:potential} one formally restores the action for General Relativity (GR) in the gauge $N=1$. It is important to emphasize, however, that 
HG and GR feature different number of propagating modes: in addition to the usual transverse-traceless tensor graviton, HG also possesses a spin-0 mode (scalar graviton)  \cite{Horava2008, Horava2009, Kobakhidze2009}. The extra degree of freedom is there due to the explicit breaking of the full diffeomorphism invariance \eqref{eq:FDiffsCoords}. As a result, in the limit $(\lambda, \eta, \mu_i, \nu_i)\to (1,1,0,0)$, HG does not recover GR but instead recovers GR+DM~\cite{Mukohyama:2009mz}, where the DM sector corresponds to the scalar graviton and behaves as dark matter. 

Quantum properties of projectable HG have been studied in detail. The model was shown to be perturbatively renormalizable in any dimensionality \cite{Barvinsky2015, Barvinsky2017Renorm}. Furthermore, in $d=3$ it has several asymptotically free UV fixed points \cite{Barvinsky2019, Barvinsky2021}, with a unique point producing regular renormalization group (RG) flow towards the infrared (IR) domain \cite{Barvinsky2023, Barvinsky2024} as conjectured earlier in \cite{Gumrukcuoglu:2011xg}.\footnote{See also \cite{DOdorico:2014tyh,DOdorico:2015pil} for non-perturbative studies of the RG; Ref. \cite{Radkovski2023} for the analysis of the observables near the fixed points, and \cite{Barvinsky2017AssFreed} for the analysis in $d=2$.} All in all, \textit{projectable} Ho\v rava gravity is perturbatively UV complete. 

So how does a UV complete $(3+1)$ dimensional QFT for the metric tensor look like at low energies?  Depending on the parameters, 
the naive perturbative expansion can remain valid or break down. In the former regime, the Minkowski background develops an infrared (IR) instability. Phenomenological constraints then require that the instability must be slow enough to be hidden by other time-dependent processes such as the Hubble expansion of the universe and the standard Jeans instability \cite{Mukohyama2010} (see (\ref{eqn:scenario2}) below). This inevitably brings the system to the latter regime where the naive perturbative expansion is invalidated \cite{Koyama2009}.
It is important to emphasize that this is not a breakdown of the theory since, even if strongly coupled, the system is still described by a finite number of parameters, thanks to the renormalizability. It has been proposed \cite{Mukohyama2010,Gumrukcuoglu:2011ef}, 
based on the fully nonlinear analysis for a spherically symmetric static vacuum configuration and the gradient expansion for nonlinear cosmological perturbations, that a resummation of non-linearities can actually suppress the interactions and make the limit $\lambda\to 1+0$ smooth, similar to the Vainshtein mechanism in massive gravity \cite{Vainshtein:1972sx}. In this limit the scalar graviton is expected to manifest itself as DM mentioned above. A self-consistent quantum analysis of this scenario is, however, challenging, see e.g. the discussion in \cite{Blas2010}. (See Sec.~\ref{sec:Outlook} for more on this point.)

The purpose of this paper is to examine, if there is another phenomenologically viable scenario without imposing the above mentioned constraint to tame the infrared instability. The property that the instability appears only in a finite range of wavenumbers and is cut at large momenta by the higher derivative terms in the potential \eqref{eq:potential} resembles the situation with Lifshitz phase transitions to a modulated phase in condensed matter systems
\cite{Lifshitz1941PhaseTransitions,HORNREICH1980387,Chaikin_Lubensky_1995}. In the latter case, the instability leads to development of static inhomogeneous profile of the order parameter which periodically depends on a spatial coordinate and thereby breaks the spatial translations down to a discrete subgroup. In this paper we explore if similar static spatially periodic solutions exist in projectable HG.

The paper is organized as follows. In Sec.~\ref{sec:Horava_Gravity} we review the IR instability of projectable HG, define the relevant parameter range and discuss our approximations. Static solutions are studied in Sec.~\ref{sec:StaticSol}. We start in Sec.~\ref{subsec:maxsym} with a classification of all maximally symmetric (homogeneous and isotropic) solutions. In Sec.~\ref{subsec:planar} we turn to solutions with planar symmetry. We prove the absence of any Lifshitz-type (quasi-)periodic solutions and show that the only solutions with planar symmetry which can potentially be stable are a subclass of the maximally symmetric solutions found earlier, written in a different slicing. 
We conclude in Sec. \ref{sec:Outlook} with a summary and discussion.
Appendix~\ref{app:Appendix} considers a complementary parameter choice with different stability properties of the solutions. Appendix~\ref{app:perts} contains analysis of scalar perturbations on the maximally symmetric spaces. 
\section{Infrared instability}
\label{sec:Horava_Gravity}
Analysis of linear perturbations around the flat background in projectable HG uncovers two kinds of modes with transverse-traceless and scalar polarizations and the dispersion relations (see e.g. \cite{Barvinsky2023Review,Herrero-Valea2023}),
\bseq
\label{eq:disprel}
\begin{align}
\label{eq:dispreltt}
    \omega^2_{tt} &= \eta k^2 + \mu_2 k^4 + \nu_5 k^6 \, , \\
    \label{eq:disprels}
    \omega^2_{s} &= \frac{1-\lambda}{1-3\lambda} \left( -\eta k^2 + \bar{\mu} k^4 + \bar{\nu} k^6 \right) \;, 
\end{align}
\eseq
where
\be
\label{mubardef}
\bar{\mu} = 8 \mu_1 + 3 \mu_2 \, , \qquad \bar{\nu} = 8 \nu_4 + 3 \nu_5 \, .
\ee
Positivity of the kinetic energy requires \cite{Horava2009,Mukohyama2010} the parameter $\lambda$ to lie either to the left of $1/3$ or to the right of $1$, 
\begin{equation}
    \lambda <1/3 \, \quad \text{or } \quad \, \lambda >1 \, .
\end{equation}
This implies that the overall factor on the r.h.s (\ref{eq:disprels}) is positive. Then the $k^2$-term in the dispersion relations of the tensor and scalar gravitons have opposite signs and one of them is necessarily unstable at low enough momenta. 
Since our aim is to recover the GR phenomenology as closely as possible, we require the $tt$-graviton to be stable, imposing a constraint $\eta>0$. The instability is then located in the scalar graviton. Note that the instability is a pure IR effect and disappears at $k>k_*$, where
\be
\label{kstar}
k_*= \sqrt{\frac{\eta}{\bar\mu}}\;,
\ee
as long as $\bar\mu$ is positive. Note also that all coefficients $\mu_2$, $\bar\mu$, $\nu_5$, $\bar\nu$ can be chosen positive simultaneously. We will assume this choice in what follows.

The dispersion relations (\ref{eq:disprel}) can be viewed from two different perspectives: top-down (UV) and bottom-up (IR). Form the UV perspective, the leading contribution is the $k^6$-term, whose properties are determined by the RG flow starting at the UV fixed points. The analysis of Refs.~\cite{Barvinsky2023,Barvinsky2024} reveals existence of RG trajectories starting at an asymptotically free fixed point with\footnote{Despite the infinite value of $\lambda$, the theory is regular and weakly coupled in this regime \cite{Radkovski2023}.}  
$\l=\pm\infty$ and flowing all the way to $\l=1+0$ or $\l=1/3-0$, with $\nu_5$ and $\bar\nu$ being positive along the flow. 
The parameter region $\l>1$ is continuously connected to the GR+DM behavior \cite{Mukohyama:2009mz,Mukohyama2010} and we focus on it in the main text.
The complementary region $\l<1/3$ is considered in Appendix~\ref{app:Appendix}.

From the UV perspective, the $\eta$ and $\mu$-terms in Eqs.~(\ref{eq:disprel}) come from relevant deformations of the RG flow and their coefficients are largely arbitrary. In particular, the parameter $\eta$ can be tuned to be small (in UV units), pushing the boundary of the instability to low momenta. In this regime the $k^6$-contribution becomes negligible in the unstable region, 
\be
\label{smalleta}
\frac{\bar\nu k_*^4}{\eta}=\frac{\bar\nu\eta}{\bar{\mu}^2}\ll 1\;,
\ee
and in the analysis of the instability one can focus on the interplay between the $k^2$ and $k^4$-terms. We will adopt this simplifying assumption in the rest of the paper.

On the other hand, from the IR perspective, the coefficient $\eta$ determines the low-energy velocity of the graviton and thus must be close to the speed of light \cite{LIGOScientific2016, EmirGumrukcuoglu2017} which is conventionally set to unity. This does not contradict our previous assumption since the definition of velocity depends on the choice of units for measuring time and distance. Rescaling the spatial momentum, $k_{\rm IR}=\sqrt{\eta} k$, we get the dispersion relations in the IR variables,
\bseq
\label{eq:disprelIR}
\begin{align}
\label{eq:disprelIRtt}
    \omega^2_{tt} &= k_{\rm IR}^2 + \frac{\mu_2}{\eta^2} k_{\rm IR}^4 + \frac{\nu_5}{\eta^3} k_{\rm IR}^6 \, , \\
    \label{eq:disprelIRs}
    \omega^2_{s} &= \frac{1-\lambda}{1-3\lambda} \left( -k_{\rm IR}^2 + \frac{\bar{\mu}}{\eta^2} k_{\rm IR}^4 + \frac{\bar{\nu}}{\eta^3} k_{\rm IR}^6 \right) \;.
\end{align}
\eseq
In these units, the speed of light is supposed to be close to one. Note that the gravitational constant also gets rescaled, so that its low-energy value is $G_{\rm IR}=G/\eta^{3/2}$. This low-energy gravitational constant sets the physical Planck mass, $M_{\rm Pl}=G_{\rm IR}^{-1/2}$.
The terms with the higher powers of momenta set the scale of breaking of the Lorentz invariance. The assumption (\ref{smalleta}) implies that the scales suppressing the quartic and sextic terms are different (assuming $\mu_2\sim\bar\mu$, $\nu_5\sim \bar\nu$),
\begin{equation}
\label{eq:scaling}
    \frac{M_{{\rm LV},4}}{M_{\rm Pl}}\sim \frac{\eta^{1/4}\sqrt{G}}{\sqrt{\bar\mu}}\;,\qquad \frac{M_{{\rm LV},6}}{M_{\rm Pl}}\sim\frac{\sqrt{G}}{\bar\nu^{1/4}}\;, \qquad  M_{{\rm LV},4}\ll M_{{\rm LV},6}\; .
\end{equation}
In other words, from the IR perspective, the assumption (\ref{smalleta}) is just stating that the Lorentz violating terms with four derivatives are enhanced compared to the terms with six derivatives.
This hierarchy is compatible with phenomenological constraints which are very broad,
\be
\label{pheno}
\frac{1}{0.01\,{\rm mm}} \lesssim M_{{\rm LV},4}\,,~M_{{\rm LV},6} \lesssim M_{\rm Pl}\;.
\ee
Here the upper bound comes from the requirement that the theory is weakly coupled at Planckian energy \cite{Horava2009,Blas2010,Blas:2009ck}, whereas the lower bound comes from the constraints of deviations from Newton's law \cite{Blas:2014aca}.\footnote{The lower bound on the scale suppressing Lorentz violation in the matter sector is much more stringent \cite{Liberati2013}.  The two scales, however, need not be the same \cite{Pospelov2010,Bednik2013,Coates:2018vub}.} 

The length scale of the scalar instability in the IR units is set by 
\be
\label{kstarIR}
k_{{\rm IR}*}=\frac{\eta}{\sqrt{\bar \mu}}\sim M_{{\rm LV},4}\;,
\ee
and thus is microscopic. On the other hand, the time scale of its growth is 
\begin{equation}
\label{eq:frequencyscale}
 \tau_* \sim \sqrt{\frac{1-3\lambda}{1-\lambda}} \, M_{{\rm LV},4}^{-1}\;,
\end{equation}
and, since $\lambda$ is attracted to $1+0$ by the RG flow, is parametrically longer than its length scale. 
This opens two directions for taming the instability. If $\tau_*$ is sufficiently longer than the time scales of other physically relevant dynamical processes, such as the Hubble expansion of the universe and the Jeans instability of inhomogeneities, then the instability of the scalar graviton is not noticeable. 
This consideration leads to the following phenomenological constraint~\cite{Mukohyama2010}. 
\begin{equation}
 0 < \frac{\lambda - 1}{3\lambda - 1} < {\rm max}\left[\frac{H^2}{k^2_{\rm IR}}\,,\ |\Phi|\right] \qquad \mbox{ for }\qquad H < k_{\rm IR} < \frac{1}{0.01\,{\rm mm}}\,,
\label{eqn:scenario2}
\end{equation}
where $H$ is the Hubble expansion rate and $\Phi$ is the Newton potential. The upper bound on the considered wavenumbers corresponds to the range probed by the experimental tests of gravity, cf.~Eq.~(\ref{pheno}).
A challenge in this scenario is the breakdown of the naive perturbative expansion \cite{Koyama2009}. A promising resummation procedure to deal with this problem was proposed in \cite{Mukohyama2010,Gumrukcuoglu:2011ef}, but its full implementation accounting for the quantum dynamics is still pending.  (See Sec.~\ref{sec:Outlook} for further discussion on this point.)

Another option is not to assume $\l$ to be too close to $1$, so that a straightforward perturbative expansion works in IR and one does not need to worry about non-perturbative dynamics nor quantum corrections. Then the instability of flat spacetime cannot be prevented, but one may wonder if it simply drives the system to a different stable solution with the characteristic spatial scale set by $k_{{\rm IR}*}^{-1}$. Indeed, we find in Sec.~\ref{subsec:maxsym} several families of both positively and negatively curved maximally symmetric solutions with constant curvature $R\sim k_{{\rm IR}*}^2$. None of these spacetimes, however, appear to provide possible end-points of the Minkowski instability. Furthermore, their curvature is too large to be phenomenologically acceptable.

An appealing way out would be to have a solution with spatial modulation at scales $k_{{\rm IR}*}^{-1}$ that on average would have zero curvature. Since the instability is inherently in the scalar mode, one could hope to still have a solution with large amount of symmetry. That would be reminiscent of the spatially modulated ground state akin to Lifshitz phases in condensed matter system \cite{Lifshitz1941PhaseTransitions} (see also \cite{Chaikin_Lubensky_1995, HORNREICH1980387} for textbook treatment). From this perspective, the assumption (\ref{smalleta}) becomes analogous to the tuning of parameters for achieving a phase transition. In this work we consider the simplest possibility and search for solutions with spatial modulation in one dimension and planar symmetry in the other two. We find, however, that such solutions do not exist and prove in Sec.~\ref{subsec:planar} a corresponding no-go theorem.

\section{Static vacuum solutions}
\label{sec:StaticSol}
In this section we discuss static solutions of projectable HG with high symmetry. We start with 
maximally symmetric spaces 
having spatial geometry of a sphere $S^3$ or a hyperboloid $H^3$. We then look for solutions possessing only a planar symmetry in the $(x,y)$-plane.
Since we are interested in static (as opposed to stationary) solutions, we impose the symmetry under time reversal, $t\mapsto -t$, which implies that the shift must vanish, $N^i = 0$. The momentum constraints obtained by variation of the action with respect to $N^i$ are then satisfied identically. Variation of Eq.~(\ref{eq:HG_Action}) with respect to the lapse then provides a global Hamiltonian constraint,  
\begin{equation}
\label{eq:Hamiltonian_Constraint}
    \int d^3 x \sqrt{\gamma}\,\mathcal{V}= 0 \, .
\end{equation}
We will consider solutions with and without imposing this constraint. Discarding it can be justified in two independent ways. First, one could formulate the theory without the lapse to start with. Indeed, as discussed above, one can always fix the gauge \eqref{eq:Projectability2}, and so locally the theory is the same either with or without the lapse. Second, we can always think of the candidate solutions as providing description of only an observable part of the Universe. The Hamiltonian constraint then can be satisfied on average over many patches~\cite{Mukohyama:2009mz} and/or different connected pieces~\cite{Matsui:2021yte}. 
Throughout this section we work in the UV units, in which the action (\ref{eq:HG_Action}) is formulated, and neglect the terms with six derivatives in the potential.
\subsection{Maximally symmetric spaces}
\label{subsec:maxsym}
We look for the metric in the form
\begin{equation}
\label{eq:ds}
    ds^2 = -dt^2 + \gamma_{ij} dx^i dx^j \, , 
\end{equation}
with maximally symmetric spatial slices and allow for a global time dependence,
\begin{equation}
    \gamma_{ij} = a^2(t) \bar{\gamma}_{ij} \quad \text{ with } \quad R_{ij} = \frac{\bar{R}}{3 a^2} \gamma_{ij} \implies R = \frac{\bar{R}}{a^2} \, .
\end{equation}
Here $\bar{\gamma}_{ij}$ is the metric in a unit sphere $S^3$ (hyperboloid $H^3$) with the curvature 
$\bar{R}=6$ ($\bar{R}=-6$).
Admitting time dependence is convenient for the study of the stability of the solutions under global perturbations. To derive the equations of motion, we substitute the Ansatz (\ref{eq:ds}) into the action (\ref{eq:HG_Action}) and obtain, up to an overall volume integral, 
\begin{equation}
\label{eq:action_superspace}
    S\propto  3(1-3\lambda) a\dot{a}^2 - V(a) \, ,
\end{equation}
with an effective potential 
\begin{eqnarray}
\label{effpot}
    V(a) = 2 \Lambda a^3 - \eta \bar{R} a + \frac{\mu \bar{R}^2}{3 a} \,.
\end{eqnarray}
Static solutions correspond to the extrema of $V(a)$.
Here we have introduced the combination
\be
\label{mudef}
\mu=3\mu_1+\mu_2\;.
\ee
The requirement  $\mu_2,\bar{\mu}>0$ for the stability of the high-energy modes implies 
\begin{eqnarray}
    \mu > - \frac{\mu_2}{8} \, ,
\end{eqnarray}
which generally allows for any  sign of $\mu$. 

Note that the kinetic term for the scale factor is negative for $\lambda>1$.
This property also holds in GR \cite{Gibbons:1978ac}. However, in GR the scale factor does not represent an independent degree of freedom due to the {\it local} Hamiltonian constraint. In HG the situation is different. If we discard the global Hamiltonian constraint for the reasons mentioned above, the scale factor becomes dynamical. Its behavior is then opposite to that of a normal particle: it will evolve away from a static solution $a_0$ if it corresponds to a {\it minimum} of $V(a)$ and will oscillate around $a_0$ if $V(a_0)$ is a local {\it maximum}. In other words, candidate stable solutions of the theory correspond to local maxima of $V(a)$.\footnote{This is a necessary but not sufficient condition for stability. A complete analysis of scalar perturbations, including inhomogeneous modes, is carried out in Appendix~\ref{app:perts}.} Note that the negative kinetic term of the homogeneous mode does not imply any rapid ghost instabilities, since the kinetic energy is positive for perturbations with short wavelengths \cite{Horava2009,Mukohyama2010} (see also Appendix~\ref{app:perts}). 

Equating the derivative of the effective potential (\ref{effpot}) to zero, we find that the static solutions satisfy
\begin{equation}
\label{eq:GenSolMaxSym}
a^2= \frac{2\bar R}{3 \eta \kappa_{\pm}} \, ,
\end{equation}
where we have introduced the combinations 
\begin{equation}
\label{eq:kappa}
     \kappa_{\pm} = \frac{\pm \sqrt{1+\mu \, \sigma}-1}{\mu}  \, , \qquad \sigma \equiv \frac{8\Lambda}{\eta^2}\;.
\end{equation}
Note that the sign of the $\mu$-independent parameter $\sigma$ coincides with the sign of the cosmological constant $\Lambda$. A pair of solutions exists iff $\mu\sigma>-1$.
Their spatial curvature reads
\begin{equation}
\label{curv}
    R =\frac{3}{2}\eta \kappa_{\pm} \; ,
\end{equation}
and coincides with the 4-dimensional spacetime curvature, because of vanishing $K_{ij}$.
As anticipated in Sec.~\ref{sec:Horava_Gravity}, the curvature is of order $k_*^2$, so these spacetimes are highly curved.

\begin{figure}[t]
    \centering
    \includegraphics[width=0.7\linewidth]{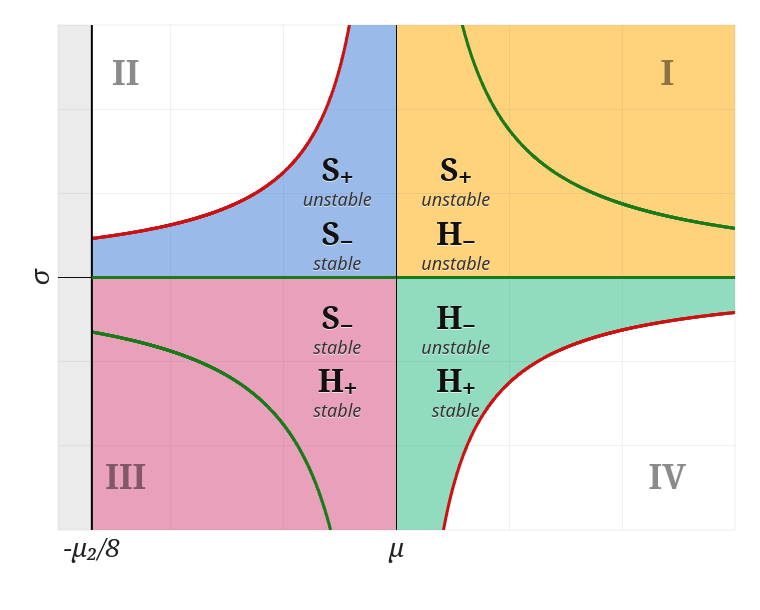}
    \caption{Static homogeneous isotropic solutions of projectable Ho\v rava gravity existing in different regions in the space of parameters $\mu=3 \mu_1 + \mu_2$ and $\sigma = 8 \Lambda/\eta^2$. Labels $S_\pm$ and $H_\pm$ stand for solutions with spatial topology of spheres $S^3$ or hyperboloids $H^3$, respectively. For each solution we indicate whether it is stable or unstable with respect to homogeneous perturbations when the kinetic coupling $\l$ is greater than $1$.
    No solutions exist at $\mu\sigma<-1$. Gray shading shows the region $\mu<-\mu_2/8$ incompatible with the stability of the high-frequency modes. Green lines mark the values of parameters, for which the solutions $S_+$ or $H_+$ satisfy the global Hamiltonian constraint. }
    \label{fig:MaxSym}
\end{figure}

The sign of the curvature is determined by the sign of $\kappa_{\pm}$ (recall that $\eta>0$). Solutions with $S^3$ ($H^3$) 
slicing exist if $\kappa_+$ or $\kappa_-$ is positive (negative). We will denote the corresponding branches of solutions as $S_{\pm}$ ($H_{\pm}$). Depending on the signs of the parameters $\mu$ and $\sigma$ we have solutions with different topologies. This is illustrated in Fig.~\ref{fig:MaxSym} which shows the pairs of solutions existing in different regions of the parameter space $(\mu, \sigma)$. 

The stability (under the global perturbations) can be established by looking at the second derivative of the effective potential $V''(a)$. As discussed above, stable solutions for $\l>1$ correspond to the maxima of the effective potential. Using the expression (\ref{eq:GenSolMaxSym}) for the scale factor, the stability condition takes the form,
\be
\label{stabcond}
\sigma+\mu\kappa_\pm^2<0\;.
\ee
It implies that the solutions $S_+$ and $H_-$ are unstable at all values of $\mu$ and $\sigma$ where they exist, whereas the solutions $S_-$, $H_+$ are stable.\footnote{The analysis of Appendix~\ref{app:perts} shows that the solution $H_+$ can be unstable with respect to {\it inhomogeneous} perturbations in a finite range of wavelengths. \label{foot5}}

Generally the solutions do not obey the global Hamiltonian constraint, unless we tune the parameters to achieve $V(a)=0$ on the solution. This can be satisfied only for the $S_+$ and $H_+$ branches under the condition 
\be
\label{Hamcond}
\mu\sigma=0\qquad \text{or}\qquad \mu\sigma=3\;.
\ee
This condition is analogous to the tuning of the cosmological constant $\Lambda$ in GR. 
In HG the solutions not satisfying the global Hamiltonian constraint are supported by the dark matter as integration constant~\cite{Mukohyama:2009mz} (DM mentioned in Sec.~\ref{eq:introduction} and Sec.~\ref{sec:Horava_Gravity}.), and they are in this sense similar to the Einstein static universe in GR which is supported by dust fluid.
The properties of the static maximally symmetric solutions found in this section are summarized in Table~\ref{tab:max}. We observe that the only branch of solutions that is stable with respect to the homogeneous perturbations and can satisfy the global Hamiltonian constraint is $H_+$.\footnote{See footnote \ref{foot5}.} 

\begin{table}[h]
\begin{center}
\begin{tabular}{|c|c|c|c|c|}
\hline
Solution & Spatial topology & Stable? & Hamiltonian constraint?  \\
\hline
$S_+$ & sphere & no & compatible \\
\hline
$S_-$ & sphere & yes & not compatible \\
\hline
$H_+$ & hyperboloid & yes & compatible \\
\hline
$H_-$ & hyperboloid & no & not compatible\\
\hline
\end{tabular}
\end{center}
\caption{Properties of static maximally symmetric solutions in projectable Ho\v rava gravity.}
\label{tab:max}
\end{table}

At $\sigma=0$ (vanishing cosmological constant) the solutions $S_+$ and $H_+$ turn into a Minkowski spacetime with flat spatial slicing. The second solution at the same values of parameters is $S_-$ for $\mu<0$ and $H_-$ for $\mu>0$. None of these solutions can be an endpoint of the Minkowski spacetime instability: the $H_-$ solution is itself unstable, whereas the solution $S_-$ has compact spatial slices and the initial non-compact flat foliation cannot evolve into it without creating singularities. We now turn to another logical possibility --- an inhomogeneous (quasi-) periodic geometry.
\subsection{Inhomogeneous geometries with planar symmetry: a no-go theorem}
\label{subsec:planar}
As discussed in Sec.~\ref{sec:Horava_Gravity},
the property that the instability of the flat space develops at non-vanishing momenta suggests that it may evolve into an inhomogeneous static space whose curvature is periodically modulated in one dimension. We now show that this expectation is not realized.

We consider a metric Ansatz with planar symmetry $ISO(2)$ in the $xy$-directions,
\begin{equation}
\label{eq:metric_ansatz}
    \gamma_{ij}dx^idx^j=e^{f(z)} \big(dx^2 + dy^2\big) + e^{g(z)} dz^2 \, ,
\end{equation}
with arbitrary time-independent functions $f(z)$ and $g(z)$. The function $g(z)$ can be gauged away by redefining the $z$-coordinate, but we keep the most general Ansatz at this point since we want to derive the equation on the metric (\ref{eq:metric_ansatz}) using the variational principle. The components of the Ricci tensor constructed from (\ref{eq:metric_ansatz}) read,
\begin{eqnarray}
    R_{11} = R_{22} = \frac{e^{f-g}}{4} \left(-2f'^2 + f' g' - 2 f'' \right) \, , \qquad R_{33} = \frac{1}{2} \left(-f'^2 + f' g' - 2 f'' \right)\;,
\end{eqnarray}
with all off-diagonal components vanishing. 
Evaluating further the scalar curvature and inserting it into the potential term, truncated to the first line in (\ref{eq:potential}), we arrive at 
\be
\begin{split}
\sqrt{\gamma} {\cal V}=\e^{f+\frac{g}{2}} \bigg[2\Lambda +\eta\e^{-g}\bigg(\frac{3f'^2}{2}-f'g'+2f''\bigg)
&+\mu \e^{-2g}\bigg(\frac{3f'^4}{4}-f'^3g'+2f'^2f''\bigg)\\
&+\bar\mu \e^{-2g}\bigg(\frac{f'^2g'^2}{8}-\frac{f'f''g'}{2}+\frac{f''^2}{2}\bigg)
\bigg]\;,
\end{split}
\ee
where $\mu$ and $\bar\mu$ are defined in Eqs.~(\ref{mudef}), (\ref{mubardef}). Taking variation with respect to $g$ and $f$ and then setting $g(z)=0$, we obtain a pair of equations,
\begin{subequations}
\label{eq:Y_eoms}
\begin{gather}
\label{eq:Y_eom1}
      E_{\Lambda}[Y] \equiv  2\bar\mu Y''Y -\bar\mu Y'^2 + 2\bar\mu Y'Y^2 - \frac{\mu}{2}Y^4 + \eta Y^2 + 4 \Lambda = 0 \;,\\
\label{eq:Y_eom2}
    E[Y] \equiv \bar\mu Y''' + \bar\mu Y''Y+2\bar\mu Y'^2 - \mu Y' Y^2 +\eta  Y'  = 0 \; ,
\end{gather}
\end{subequations}
where we have introduced the notation $Y(z) \equiv f'(z)$.
Note that the two equations \eqref{eq:Y_eoms} are not independent: they are related by
\begin{equation}
\label{eq:eom_relation}
    E[Y] =  \frac{1}{2 Y} \partial_z E_{\Lambda}[Y] \; .
\end{equation}

We make a key observation. The equation \eqref{eq:Y_eom2} can be written in the form
\begin{equation}
\label{eq:Lyapunov}
    E[Y] = 0 \quad \implies \quad \partial_zF[Y] = - \bar\mu Y'^2  \, , 
\end{equation}
with the function
\begin{equation}
    F[Y]\equiv \bar\mu Y'' + \bar\mu Y Y'  - \frac{\mu}{3} Y^3 + \eta Y\, .
\end{equation}
This immediately implies that there can be no non-trivial solutions periodic in $z$. Indeed, for a periodic solution $Y$, the function $F$ must be periodic as well. On the other hand, according to \eqref{eq:Lyapunov}, $F$ is monotonically decreasing unless $Y'$ identically vanishes. But the solution with constant $Y=Y_0$ corresponds to the linear metric function
\begin{equation}
\label{eq:fHyper}
    f(z) = Y_0 z + \text{const} \, ,
\end{equation}
and thus is not periodic either. We conclude that Eqs.~(\ref{eq:Y_eoms}) do not have any non-trivial periodic solutions. 

We can further extend this no-go result and classify all bounded solutions, i.e. solutions with $Y(\pm \infty) < \infty$ and finite values of derivatives. Integrating \eqref{eq:Lyapunov} we observe that for a bounded solution the derivative of $Y$ must vanish at $z=\pm \infty$, and thus all bounded solutions must asymptote to those with $Y(z)=\text{const}$. 
From Eq.~\eqref{eq:Y_eom1} restricted to a constant $Y=Y_0$ we find possible solutions,
\begin{equation}
\label{eq:roots}
    Y_0^2 = -\eta \kappa_{\pm}\;,
\end{equation}
with $\kappa_\pm$ from (\ref{eq:kappa}). Clearly, they exist only in the regions of the parameter space where at least one of $\kappa_{+}$ or $\kappa_-$ is negative. These are the regions I, III and IV in Fig.~\ref{fig:MaxSym}. The solutions have constant negative curvature given by Eq.~(\ref{curv}), which suggests that they are nothing but the maximally symmetric spacetimes with hyperboloid spatial slices $H_{\pm}$ found in Sec.~\ref{subsec:maxsym}. This is indeed the case, as can be seen explicitly by performing the change of coordinates
\begin{equation}
    u = \frac{2}{Y_0} \, e^{-Y_0 z/2}\;,
\end{equation}
which brings the spatial metric to the form
\begin{equation}
    \gamma_{ij}dx^idx^j = \frac{4}{Y_0^2} \frac{dx^2 + dy^2 + du^2}{u^2} \, .
\end{equation}
This is the metric of a hyperboloid in Poincar\'e coordinates.

Are there any static planar bounded solutions with non-constant $Y(z)$?
Up to now, we have established that any such solution must approach at $z\to\pm\infty$ one of the hyperbolic spaces $H_\pm$. As discussed in Sec.~\ref{subsec:maxsym}, the branch $H_-$ is unstable (at $\l>1$), and the solutions with these asymptotics are not of interest to us here.\footnote{These solutions may be stable for $\l<1/3$ and are considered in Appendix~\ref{app:Appendix}.} Thus we focus on geometries with the asymptotics $H_+$. The corresponding values of $Y_0$ are 
\begin{equation}
\label{eq:Yzerokappaplus}
    Y_0 = \pm \sqrt{\eta |\kappa_+|} \, .
\end{equation}
Equation (\ref{eq:Lyapunov}) forbids any solutions with $Y'\neq 0$ and the same value of $Y$ on both infinities. Further, it requires
\be
\label{Fineq}
\left(\frac{\mu\kappa_+}{3}+1\right)Y(-\infty)>\left(\frac{\mu\kappa_+}{3}+1\right)Y(\infty) \, ,
\ee
and since $(\mu\kappa_+/3+1)>0$, the only possibility is that a solution
interpolates between ${Y(-\infty)=\sqrt{\eta |\kappa_+|}}$ and $Y(\infty)=-\sqrt{\eta |\kappa_+|}$. 
Such a solution would have to cross $Y=0$ at some point $z_0$. At this point Eq.~(\ref{eq:Y_eom1}) would take the form,
\begin{eqnarray}
\label{nogoeq}
   -\bar\mu Y'^2 +4 \Lambda = 0\, .
\end{eqnarray}
This, however, is impossible to satisfy since the $H_+$ solutions exist only for negative values of the cosmological constant $\Lambda$, see Fig.~\ref{fig:MaxSym}. 
We conclude that there are no non-trivial static solutions with planar symmetry interpolating between two $H_+$ asymptotics.

\section{Summary and Discussion}
\label{sec:Outlook}

In this paper we explored the low-energy properties of projectable HG. Since the Minkowski spacetime in this theory is known to be unstable against long-scale perturbations, the spacetime representing our universe needs to be away from Minkowski, exhibiting either time-dependence or space-dependence. The time and length scales associated with the two respective scenarios could be quite different\footnote{From the dispersion relation of the scalar graviton, the ratio between them is expected to be of order $(\lambda-1)^{-1/2} \to\infty$ in the IR limit $\lambda\to 1+0$.} and, as a result, the outlook for the two scenarios could also differ significantly. In the former scenario, the 
coupling $\lambda$ must be attracted sufficiently close to $1+0$ in the IR by the RG flow 
so that the instability is hidden by time-dependent processes such as the Hubble expansion of the universe and the standard Jeans instability. In the latter scenario, on the other hand, one would expect existence of alternative ground states.

We have  performed a search for such ground states. Since the instability occurs for perturbations in a finite range of wavenumbers, we speculated that the ground state could have a (quasi-)periodic dependence on a space coordinate, similar to a modulated phase in magnetic condensed matter systems. The fact that the instability is exhibited by the scalar graviton mode motivated a metric Ansatz with planar symmetry $ISO(2)$ in the orthogonal directions. To simplify the analysis, we truncated the Lagrangian to terms with four or fewer spatial derivatives. The four-derivative terms are sufficient to cut the instability at high momenta, thereby ensuring good UV properties of the perturbations. We kept the cosmological constant $\Lambda$ as a free parameter to explore its possible role in stabilizing the solutions.  

Our main result is negative: we have proved that no static planar solutions with quasi-periodic metric exist. Generally, the solutions with planar symmetry can be only of two types. The solutions of the first type have enhanced symmetry $SO(3,1)$ and correspond to spacetimes with hyperbolic spatial slices. We found two branches of such solutions which we called $H_\pm$. Depending on the parameter choice, none, one or both can exist, see Fig.~\ref{fig:MaxSym}. Their stability properties depend on the value of the kinetic coupling $\l$. For the phenomenologically interesting choice $\l\gtrsim 1$, the solution $H_-$ is unstable w.r.t. homogeneous perturbations of the scale factor, whereas the solution $H_+$ is stable against such perturbations. However, for small values of the cosmological constant, the solution $H_+$ reduces to the Minkowski space, so at least for these values of $\Lambda$, it is unstable w.r.t. perturbations with finite wavenumbers. For another choice of $\l$ compatible with unitarity, $\l<1/3$, the situation for the solution $H_-$ gets reversed: as we show in Appendix~\ref{app:perts}, it is stable against all scalar perturbations, and thus likely stable altogether.\footnote{We do not expect any instability from the tensor modes, because these are stable even around the Minkowski space.} 
 
The second possible type of static planar solutions describe domain walls interpolating between two hyperbolic geometries. In each case $\l>1$ or $\l<1/3$ we studied geometries with potentially stable asymptotics $H_+$ or $H_-$, respectively. In the first case, we showed the absence of any domain walls connecting two $H_+$ regions. In the second case, domain walls between two $H_-$ regions do exist for special values of parameters (see Appendix~\ref{app:A2}). We have not studied their stability.

For completeness, we also studied other maximally symmetric solutions which we found to have compact spherical spatial slices. We called two branches of such solutions $S_\pm$ and analyzed their stability with the following results. For $\l>1$: $S_+$ is unstable, whereas $S_-$ is stable. For $\l<1/3$: $S_+$ can be stable or not, depending on the parameter choice, whereas $S_-$ is unstable.

None of the solutions we found in this work can serve as end-points of the development of the Minkowski instability in the phenomenologically interesting region $\l\gtrsim 1$. This strongly suggests that the instability leads to a time-dependent spacetime, never reaching any ground state. 

Let us discuss possible caveats to this conclusion. Our no-go results pertain to metrics with maximal and planar symmetries and do not exclude putative ground states with a lower symmetry, or no symmetries at all. While such possibilities are hard to rule out, we believe them to be unlikely. Another limitation of our analysis is the neglect of the terms with six spatial derivatives which are present in the full HG Lagrangian. We do not expect these terms to qualitatively change our results, but their inclusion would be required to fully rule out inhomogeneous ground states. Finally, in our analysis we took the coupling $\eta$ in the potential (\ref{eq:potential}) to be positive, which leads to the instability in the scalar graviton sector. In principle, one can shift the instability from scalar to tensor graviton by choosing $\eta<0$, and ask if a static ground state exists in this case. This option, however, appears to be of purely academic interest since it is hard to see how it could be compatible with the known low-energy properties of the gravitational waves. 

In summary, our result shows difficulties in taming the infrared instability of projectable HG by space-dependence alone. Let us thus go back to the first scenario relying on time-dependence to reconcile projectable HG with observations and discuss the required time-dependence at short and long spatial scales. 

At short spatial scale ($\ll 0.01\mathrm{mm}$), projectable HG might develop a kind of spacetime foam: strong curvature inhomogeneities on short length and time scales inaccessible to present-day experiments, which on long distances and times average out to produce a smooth geometry. This picture assumes the absence of caustics and the slowness of cascade from short to long distance scales. While the first step towards the absence of caustics was provided in \cite{Mukohyama:2009tp}, the slowness of cascade from short to long distance scales is extremely hard to assess by analytical methods and its exploration would require full-fledged dynamical numerical simulations. 

At long spatial scale ($\gg 0.01\mathrm{mm}$), on the other hand, one may entertain the possibility that the instability is sufficiently slow to be hidden behind other physically relevant time-dependent processes such as the Hubble expansion of the universe and the Jeans instability of inhomogeneities.  This scenario 
requires the effective coupling $\lambda$ obtained upon averaging over short-scale fluctuations to be 
very close to $1+0$ as in (\ref{eqn:scenario2}), which
invalidates the naive perturbative expansion in IR and challenges the applicability of the classical description based on the action (\ref{eq:HG_Action}). Nevertheless, the fully nonlinear analysis of spherically symmetric static configurations~\cite{Mukohyama2010} and the gradient expansion for nonlinear cosmological perturbations~\cite{Gumrukcuoglu:2011ef} show continuity in the $\l\to1+0$ limit, with the scalar graviton behaving as a built-in dark matter component. The interactions of the dark-matter-like component with the rest of the system including ordinary matter and with itself remain weak in the limit, showing no sign of strong coupling at the classical level. This suggests that there might be a change of variables in the scalar graviton sector that would keep the perturbative expansion under control. While the analysis of \cite{Mukohyama2010,Gumrukcuoglu:2011ef} is classical, it also points a promising direction towards consistent quantum treatment. After solving the momentum constraint order by order in perturbations, each term with negative powers of $(\lambda -1)$ involves precisely two time derivatives, this structure stemming from the spatial diff-invariance and renormalizability. A proper redefinition of the scalar graviton variable to resum those diverging kinetic terms, if possible to find, should significantly ease the analysis in the $\lambda\to 1+0$ limit. It is important to seek such nonlinear redefinition of the scalar graviton to shed light on the quantum properties of projectable HG in the $(\l-1)\ll 1$ regime.

\section*{Acknowledgments}
We thank Sung-Sik Lee for insightful discussions. The work of JR and SS is supported by the 
Natural Sciences and Engineering Research Council (NSERC) of Canada. Research at Perimeter Institute is supported in part by the Government of Canada through the Department of Innovation, Science and Economic Development Canada and by the Province of Ontario through the Ministry of Colleges and Universities. The work of S. M. was supported in part by JSPS (Japan Society for the Promotion of Science) KAKENHI Grant No. JP24K07017 and World Premier International Research Center Initiative (WPI), MEXT, Japan.
\appendix

\section{Solutions at $\lambda<1/3$}
\label{app:Appendix}
In the main body of this paper we have focused on the regime $\lambda>1$. 
Here we analyze the regime of the theory with $\lambda<1/3$. In this case the kinetic terms in the action is always positive:
\be
\label{positiveLkin}
K_{ij}K^{ij}-\l K^2= \left(K_{ij}-\frac{1}{3}\gamma_{ij}K\right)^2+\left(\frac{1}{3}-\l\right) K^2>0\;.
\ee
In particular, it is positive for homogeneous perturbations of the scale factor, so that 
the solutions $S_-$ and $H_+$ are unstable. On the other hand, 
the other two families of solutions, $S_+$ and $H_-$, are stable with respect to homogeneous perturbations.
In what follows we demonstrate two properties. First, the solutions that are of the 
$S_+$ type and satisfy the Hamiltonian constraint for positive cosmological constant
provide an absolute minimum of the potential energy, implying that they are absolutely stable.\footnote{A complementary analysis in Appendix~\ref{app:perts} shows that for other values of the cosmological constant, the solution $S_+$ may or may not be destabilized by inhomogeneous scalar perturbations. The solution $H_-$ is always stable at $\l<1/3$ against such perturbations.}
Second, for some values of parameters, two solutions with $H_-$ asymptotics can be connected by a domain wall.

\subsection{Stable spheres}
\label{app:A1}
Let us choose $\mu>0$ and tune the value of the cosmological constant to satisfy the second of Eqs.~(\ref{Hamcond}), 
\be
\Lambda=\frac{3\eta^2}{8\mu}\;.
\ee 
In this way we fulfill the Hamiltonian constraint. 
Further, we can turn the potential into a perfect square,
\begin{equation}
    \mathcal{V} =  \frac{3 \eta^2}{4\mu} -\eta R + \mu_1 R^2 + \mu_2 R_{ij}^2 = \mu_2 \left[\frac{\eta}{2 \sqrt{\mu_2\mu}} \gamma_{ij} -  \Bigg(1+ \sqrt{\frac{\mu}{\mu_2}}\Bigg)\frac{R}{3}\gamma_{ij}+R_{ij}  \right]^2 \, ,
\end{equation}
where we have used the definition of $\mu$, Eq.~(\ref{mudef}). 
The potential is minimized on configurations satisfying
\begin{equation}
\label{eq:BPS_eom}
      \frac{\eta}{2 \sqrt{\mu_2\mu}} \gamma_{ij} - \frac{R}{3}\gamma_{ij} \left(1+ \sqrt{\frac{\mu}{\mu_2}}\right) +R_{ij} = 0 \, .
\end{equation}
Taking the trace we get $R = \frac{3\eta}{2 \mu}$. Plugging this back into \eqref{eq:BPS_eom} we obtain the unique solution
\begin{equation}
    R_{ij} = \frac{\eta}{2 \mu} \gamma_{ij} \, ,
\end{equation}
which has the spatial topology of a sphere $S^3$. It is straightforward to see that it is a part of the branch $S_+$. 

\subsection{Domain walls between hyperboloids}
\label{app:A2}
The discussion carried out in Sec.~\ref{subsec:planar} that led us to the conclusion that bounded solutions with planar symmetry must approach hyperbolic spatial geometries at $z\to\pm\infty$ does not depend on $\l$, so it applies also at $\l<1/3$. The difference is that now the stable solutions are $H_-$, rather than $H_+$, so that instead of Eq.~(\ref{eq:Yzerokappaplus}), we shall now consider the asymptotics
\begin{equation}
\label{eq:Yzerokappaminus}
    Y_0 = \pm \sqrt{\eta |\kappa_-|} \, .
\end{equation}
Instead of Eq.~(\ref{Fineq}) we have
\be
\label{Fineq1}
\left(\frac{\mu\kappa_-}{3}+1\right)Y(-\infty)>\left(\frac{\mu\kappa_-}{3}+1\right)Y(\infty)\;,
\ee
which, depending on the value of $\mu\sigma$, implies different ordering of the asymptotics,
\be
Y(-\infty)>Y(\infty)~~~\text{if}~~\mu\sigma<3~,\qquad\qquad
Y(-\infty)<Y(\infty)~~~\text{if}~~\mu\sigma>3~,
\ee
The solutions still need to cross $Y=0$ at some point. However, we cannot use Eq.~(\ref{nogoeq}) to prove their absence since the spacetimes $H_-$ are compatible with $\Lambda>0$ (region I in Fig.~\ref{fig:MaxSym}).

It happens that for positive $\Lambda$, corresponding to $\mu\sigma>0$, a simple domain wall Ansatz,
\be
\label{eq:wall}
Y(z)=A\tanh k z \, ,
\ee
goes through the equations
of motion (\ref{eq:Y_eoms}). Since the second equation is a consequence of the first one, we can focus on (\ref{eq:Y_eom1}). Substituting (\ref{eq:wall}) into it, we find that it will be satisfied if the following relations hold:
\bseq
\label{DWeqs}
\begin{align}
\label{DWeq1}
&\mu A^4-2\eta A^2-8\Lambda=0\;,\\
\label{DWeq2}
    &4\bar\mu k^2 - 2 \bar\mu A k -\mu A^2+ \eta = 0 \; , \\
    \label{DWeq3}
    &6\bar \mu k^2 - 4\bar\mu A k - \mu A^2 = 0 \; .
\end{align}
\eseq
From Eq.~(\ref{DWeq1}) we find $A^2=-\eta \kappa_-$, as expected.\footnote{We discard the second root $A^2=-\eta \kappa_+$ corresponding to the $H_+$ asymptotics which do not lead to any non-trivial solutions, as shown in the main text. } From the other two equations we then obtain
\begin{align}
k^2=\frac{\eta\mu}{2\bar\mu}\kappa_+~,\qquad Ak=\frac{\eta}{4} (3\kappa_++\kappa_-)\;.
\end{align}
These relations impose constraints on the theory parameters. Expressing $\kappa_+=-\kappa_--2/\mu$ and solving for $\kappa_-$, we obtain the following options:
\bseq
\begin{align}
\label{DW1}
&\mu\kappa_-=\frac{-3+2 u+\sqrt{6u+4u^2}}{1-2u}~,~~~~~u>0\;,\\
\label{DW2}
&\mu\kappa_-=\frac{-3+2 u-\sqrt{6u+4u^2}}{1-2u}~,~~~~~0<u<1/2\;,
\end{align}
\eseq
where we have introduced $u\equiv\mu/\bar\mu$. The family (\ref{DW1}) corresponds to $\mu\sigma<3$ and, according to Eq.~(\ref{Fineq1}) interpolates between $Y(-\infty)=\sqrt{\eta|\kappa_-|}$ and $Y(\infty)=-\sqrt{\eta|\kappa_-|}$. On the other hand, the family (\ref{DW2}) corresponds to $\mu\sigma>3$ and interpolates from $Y(-\infty)=-\sqrt{\eta|\kappa_-|}$ to ${Y(\infty)=\sqrt{\eta|\kappa_-|}}$. 
\section{Scalar perturbations on maximally symmetric spaces}
\label{app:perts}
In this Appendix we derive the quadratic Lagrangian for the space- and time-dependent scalar perturbations around the maximally symmetric backgrounds found in Sec.~\ref{subsec:maxsym} and analyze their stability. 
To perform the computations, we use the Mathematica \cite{Mathematica} package {\it xAct} \cite{xAct}. 

The most general scalar perturbation can be written in the form,
\begin{align}
\label{metrexpand}
    \delta\gamma_{ij} = a_0^2(2 \bar\gamma_{ij} \psi + \bar\nabla_i \bar\nabla_j E )\;, \qquad~~~ \delta N_i = a_0^2\bar\nabla_i B \, ,
\end{align}
where $\bar\gamma_{ij}$ is the metric of a unit 3-dimensional sphere or hyperboloid,
\begin{align}
\bar\gamma_{ij}dx^idx^j=\begin{cases}
d\chi^2+\sin^2\chi\,(d\theta^2+\sin^2\theta \,d\varphi^2)\;,& \quad\text{sphere}\\
d\chi^2+\sinh^2\chi\,(d\theta^2+\sin^2\theta \,d\varphi^2)\;,&\quad \text{hyperboloid}
\end{cases}
\end{align}
and $a_0$ is the scale factor of the static solution given by Eq.~(\ref{eq:GenSolMaxSym}). The covariant derivatives are taken with respect to the background metric $\bar\gamma_{ij}$. To simplify notations, we will omit the overbars on the derivatives in what follows.

Substituting the expansion (\ref{metrexpand}) into the kinetic part of the action (\ref{eq:HG_Action}) and integrating by parts to simplify the result, we obtain the quadratic kinetic Lagrangian,
\begin{align}
\label{Lag2kin}
    \sqrt{\gamma}{\cal L}^{(2)}_{\rm kin}  = a_0^3\sqrt{\bar\gamma}\bigg[&3(1-3\l) \dot{\psi}^2 - 2(1-3\l)
    \dot{\psi} \Delta \left(B-\frac{\dot{E}}{2}\right) \notag\\
    &+ (1-\l)\left(B-\frac{\dot{E}}{2}\right) \Delta^2\left(B-\frac{\dot{E}}{2}\right) + 2 \mathcal{K}\left(B-\frac{\dot{E}}{2}\right) \Delta\left(B-\frac{\dot{E}}{2}\right) \bigg],
\end{align}
where $\Delta\equiv\nabla_i\nabla^i$ is the Laplacian, and ${\cal K}=+1$ ($-1$) for sphere (hyperboloid). Variation with respect to $B$ yields the momentum constraint,
\begin{equation}
\label{momconstr}
     -(1-3\lambda) \Delta \dot{\psi}+(1-\lambda) \Delta^2 \left(B - \frac{\dot{E}}{2} \right)
          + 2 \mathcal{K}\Delta\left(B - \frac{\dot{E}}{2} \right) = 0 \, ,
\end{equation}
whence we express
\be
\label{BEexpr}
B-\frac{\dot E}{2}=\frac{1-3\l}{(1-\l)\Delta+2{\cal K}}\dot\psi\;.
\ee
Substituting back into (\ref{Lag2kin}) we obtain,
\begin{equation}
\label{Lag2kin1}
    \sqrt{\gamma}\mathcal{L}^{(2)}_{\rm kin} = a_0^3\sqrt{\bar\gamma}\bigg[ \frac{2(1-3 \lambda)}{1-\lambda}\, \dot{\psi} \frac{\Delta + 3  \mathcal{K}}{\Delta + \frac{2}{1-\lambda} \mathcal{K}} \dot{\psi} \bigg] \, .
\end{equation}
We observe that the kinetic Lagrangian depends only on the gauge invariant metric perturbation~$\psi$. 

For the potential part, we use the expansion of the Ricci tensor up to quadratic order,
\be
R_{ij}=2{\cal K}\bar\gamma_{ij}-\bar\gamma_{ij}\Delta\psi-\nabla_i\nabla_j\psi+2\bar\gamma_{ij}\psi\Delta\psi+\bar\gamma_{ij}(\nabla_k\psi)^2+
2\psi\nabla_i\nabla_j\psi+3\nabla_i\psi\nabla_j\psi+(E\text{-terms})\;,
\ee
where we do not write the terms with the $E$-field explicitly.
We substitute this expression into the first line of Eq.~(\ref{eq:potential}), integrate by parts and use the background relation (\ref{eq:GenSolMaxSym}). Upon this procedure, the $E$-dependent terms cancel, as required by gauge invariance, and we obtain, 
\begin{equation}
\label{Lag2pot}
    \sqrt{\gamma}\mathcal{L}^{(2)}_{\rm pot} = a_0 \sqrt{\bar\gamma}\bigg[- 2\eta\, \psi \big(\Delta + 3 \mathcal{K} \big)\bigg(1 + \mu \kappa_{\pm}+ \mathcal{K}\frac{\bar\mu\kappa_\pm}{4} \Delta \bigg)  \psi \bigg]\, .
\end{equation}
Let us analyze separately the modes on $S_\pm$ and $H_\pm$.

\paragraph{Spheres.}
The spectrum of Laplacian on a unite sphere $S^3$ is discrete:
\begin{equation}
\label{SphLapl}
    \Delta \psi_n = - (n^2 - 1) \psi_n \; , ~~~~~~~n=1,2,3,\ldots\;.
\end{equation}
Explicitly, the eigenmodes with purely radial dependence have simple form,
\begin{equation}
\label{Sphmodes}
    \psi_{n} (t,\chi)= \frac{\sin{n \chi}}{\sin{\chi}} \,\tilde\psi_n(t)\, .
\end{equation}
Combining Eqs.~(\ref{Lag2kin1}), (\ref{Lag2pot}) and recalling that ${\cal K}=1$ for spheres, we obtain the quadratic Lagrangian for the $n$th eigenmode: 
\be
\label{Lag2n}
\sqrt{\gamma}{\cal L}^{(2)}_n=\sqrt{\bar\gamma}\bigg[a_0^3 \frac{2(1-3 \lambda)}{1-\lambda} \, \frac{n^2-4}{n^2 - \frac{3-\lambda}{1-\lambda}}\, \dot{\tilde\psi}_{n}^2
-\eta a_0(n^2-4)\bigg(-2(1+\mu\kappa_\pm)+\frac{\bar\mu\kappa_\pm}{2}(n^2-1)\bigg)\,\tilde\psi_n^2
\bigg].
\ee
Note that the action vanishes identically for the dipole mode with $n=2$. This is consistent with this mode being a pure gauge. Indeed, this mode is purely longitudinal, 
\begin{equation}
    \psi_2(t,\chi) = 2\cos\chi \,\tilde\psi_2(t) =\nabla_\chi^2\big(-2\cos\chi \,\tilde\psi_2(t)\big)\, ,
\end{equation}
and thus has the same form as the $E$-term in (\ref{metrexpand}).

Let us focus on $\lambda>1$. The kinetic term is negative for the homogeneous mode ($n=1$), in agreement with the analysis of Sec.~\ref{subsec:maxsym}. On the other hand, it is positive for all modes with $n\geq 3$, and thus no ghost instabilities at short wavelengths arise. 

To analyze the potential term, we need to further specify the branch of solutions. For the family $S_+$, the potential energy can be positive or negative, depending on the value of $n$: it is positive for $n=1$, then becomes negative in the range 
\be
\label{S+inst}
4<n^2<1+\frac{4\sqrt{1+\mu\sigma}}{\bar\mu \kappa_+}\;,
\ee
if this range is not empty, 
and becomes positive again at larger $n$. Thus, the solution $S_+$ is unstable with respect to the homogeneous mode and the modes in the range (\ref{S+inst}), but shorter modes are stable. 
This is the same situation as in flat space and it is instructive to see how the flat-space limit arise from (\ref{Lag2n}). We scale the parameters as follows,
\be
\label{flatlimit}
a_0\to\infty~,~~~\kappa_+\to 0~,~~~~a_0^2 \kappa_+=\frac{4}{\eta}~\text{---~fixed}~,~~~n\to \infty~,~~~\frac{n}{a_0}=k~\text{---~fixed}\;,
\ee
and arrive at
\be
\label{Lag2flat}
\sqrt{\gamma}{\cal L}^{(2)}_n~\propto~  \frac{2(1-3 \lambda)}{1-\lambda} \,  \dot{\tilde\psi}_{n}^2
-2 k^2\big(-\eta+\bar\mu k^2\big)\,\tilde\psi_n^2\;.
\ee
This is precisely the Lagrangian for scalar perturbations in flat spacetime leading to the first two terms in the dispersion relation (\ref{eq:disprels}). 

For the solutions $S_-$, the potential term is negative for $n=1$ and positive for all other $n$. In other words, the kinetic and potential term change sign synchronously. Remarkably, the solutions $S_-$ turn out to be stable against any scalar perturbations for $\l>1$. 

Consider now the case $\l<1/3$. The kinetic term is then positive for all $n$, including the homogeneous mode, in agreement with Eq.~(\ref{positiveLkin}). 
The solutions $S_-$ ($S_+$) are unstable (stable) against homogeneous perturbations, as discussed in the main text. A solution $S_+$ can be stable entirely if the range (\ref{S+inst}) is empty. This happens, e.g., for the solutions $S_+$ satisfying the Hamiltonian constraint and studied in Appendix~\ref{app:A1}.

\paragraph{Hyperboloids.}
The spectrum of the Laplacian is continuous, 
\begin{equation}
\label{HypLapl}
    \Delta \psi_p = - (p^2 + 1) \psi_p \, ,
\end{equation}
and the radially symmetric eigenmodes are 
\begin{equation}
    \psi_{p}(t,\chi) = \frac{\sin{p \chi}}{\sinh{\chi}}\, \tilde\psi_p(t) \, .
\end{equation}
The modes are $\delta$-function normalizable if $p$ is real. The requirement for the modes to decrease at infinity also admits imaginary values 
\be
\label{pim}
p=iq~,\qquad 0<q\leq 1\;.
\ee
These modes, however, are not $\delta$-function normalizable since the integration measure 
$\int d\chi \sinh^2\chi$ quickly diverges at $\chi\to\infty$. Whether they should be included or not in the spectrum of perturbations depends on the precise boundary conditions at the spacetime boundary. We do not attempt to formulate such boundary conditions in this paper and only note that we believe at least the constant mode corresponding to $q=1$ is physical
since it describes the cosmological expansion of the universe as a whole.

Substituting the eigenvalues (\ref{HypLapl}) into the general expressions (\ref{Lag2kin1}), (\ref{Lag2pot}), we obtain the quadratic Lagrangian for the $p$-mode,
\be
\label{Lag2p}
\sqrt{\gamma}{\cal L}^{(2)}_p=\sqrt{\bar\gamma}\bigg[a_0^3 \frac{2(1-3 \lambda)}{1-\lambda} \, \frac{p^2+4}{p^2 + \frac{3-\lambda}{1-\lambda}}\, \dot{\tilde\psi}_{p}^2
-\eta a_0(p^2+4)\bigg(-2(1+\mu\kappa_\pm)-\frac{\bar\mu\kappa_\pm}{2}(p^2+1)\bigg)\,\tilde\psi_p^2
\bigg].
\ee
Let us first look at the kinetic term in the case $\l>1$. For $p^2=-1$ ($q=1$), we recover the negative kinetic term of the homogeneous mode. On the other hand, for $p^2\to \infty$, the kinetic term is the same as in flat space and is positive. Somewhat surprisingly, the kinetic term diverges at
\be
\label{pstar}
p_*^2=\frac{3-\l}{\l-1}\;,
\ee
which for $1<\l<3$ is positive. We do not know if this behavior signals any pathologies. At the quadratic order, it implies that 
the dispersion relation vanishes at $p_*$. In particular, the solution $H_+$, which is found to be stable against the homogeneous perturbations in the main text, happens to be unstable against the modes with finite $p$ in a range
\be
\min\bigg\{p_*^2,-1+\frac{4\sqrt{1+\mu\sigma}}{\bar\mu|\kappa_+|}\bigg\}<p^2<\max\bigg\{p_*^2,-1+\frac{4\sqrt{1+\mu\sigma}}{\bar\mu|\kappa_+|}\bigg\}.
\ee
This range is always non-empty if we include the non-normalizable modes (\ref{pim}). Otherwise, it can be empty for some choices of the model parameters.
On the other hand, the solution $H_-$ is unstable for all $p^2<p_*^2$.

If $\l<1/3$, the kinetic term is positive at all values of $p^2>-1$. Hence the stability of the solutions is solely determined by the potential term. We find that the branch $H_+$ is unstable at
\be
p^2<-1+\frac{4\sqrt{1+\mu\sigma}}{\bar\mu|\kappa_+|}\;,
\ee
whereas the branch $H_-$ is stable at all $p^2>-1$.

\printbibliography[heading=bibintoc]

\end{document}